\documentstyle[12pt]{article}
\parskip
1mm
\begin{document}
\newcommand{\la}{\langle}
 \newcommand{\ra}{\rangle}
 \newcommand{\bear}{\begin{eqnarray}}
 \newcommand{\ear}{\end{eqnarray}}
 \newcommand\noi{\noindent}
 \newcommand\bo{\boldmath}
 \newcommand{\be}{\begin{equation}}
 \newcommand{\ee}{\end{equation}}
 \newcommand\beq{\begin{equation}}
 \newcommand\eeq{\end{equation}}
 \newcommand\beqn{\begin{eqnarray}}
 \newcommand\eeqn{\end{eqnarray}}
 \newcommand{\doublespace}
 {
 \renewcommand{\baselinestretch}
 {1.6}
 \large\normalsize}
 \renewcommand{\thefootnote}{\fnsymbol{footnote}}
 \def\sel{\sigma_{el}^{VN}}
 \def\sin{\sigma_{in}^{VN}}
 \def\stot{\sigma_{tot}^{VN}}
 \def\inf{\int_{-\infty}^{\infty}}


\vspace*{1.0cm}
\begin{center}

 {\Large{\bf Bremsstrahlung of a Quark\\ Propagating through a Nucleus}}
\end{center}
 
\vspace{.5cm}

\begin{center}
 {\large
Boris~Z.~Kopeliovich$^{1,2}$, Andreas Sch\"afer$^{4}$\\

\medskip

and Alexander~V.~Tarasov$^{2,3}$}
\medskip

{\sl
$^1$
Max-Planck
Institut
f\"ur
Kernphysik,
Postfach
103980,
69029
Heidelberg}

{\sl
$^2$
Joint
Institute
for
Nuclear
Research,
Dubna,
141980
Moscow
Region}

{\sl $^3$Institut
f\"ur
Theoretische
Physik
der
Universit\"at,
Philosophenweg
19, 69120
Heidelberg}

{\sl $^4$Institut f\"ur Theoretische Physik,
Universit\"at Regensburg,
93040 Regensburg}
\\

\end{center}

\vspace{.5cm}
\doublespace
\begin{abstract}
The density of gluons produced in the central rapidity 
region of a heavy ion collision is poorly known.
We investigate the influence of the effects of quantum coherence
on the transverse momentum distribution
of photons and gluons radiated by a quark propagating through
nuclear matter. 
We describe the case that the
radiation time substantially exceeds the nuclear radius
(the relevant case for RHIC and LHC energies), which is different
from what is known as Landau-Pomeranchuk-Migdal effect corresponding
to an infinite medium. 
We find {\it suppression} of the radiation spectrum at small
transverse photon/gluon momentum $k_T$,
but {\it enhancement} for $k_T>1\,$GeV.
Any nuclear effects vanish
for $k_T  \geq 10\,$GeV. Our results  
allow also to calculate the $k_T$ dependent nuclear effects 
in prompt photon, light and heavy (Drell-Yan)
dilepton and hadron production.

\end{abstract}

\bigskip


\newpage

\doublespace

\section{Introduction}

One of the major theoretical problems in  relativistic
heavy ion physics is the reliable calculation of 
gluon bremsstrahlung in the central rapidity region.
It is one of the determining factors for the general
dynamics of heavy-ion collisions, the approach to thermodynamic equilibrium 
and the possible formation of a quark-gluon plasma-like state.
This problem has been approached by a variety of ways.
We do not want to discuss the relative draw-backs and 
merits of the various approaches here and we will only cite those, which are
directly related to ours.

In this paper we consider bremsstrahlung
of photons and gluons resulting from the interaction of
a projectile quark with a nucleus for the case  that the radiation 
time is much longer than the time needed to cross the nucleus. 
This radiation or formation time was introduced in \cite{lp}
and can be presented as,
\beq
t_f = \frac{{\rm cosh}\,y}{k_T}\approx
\frac{2\omega}{k_T^2}\ ,
\label{1.1}
\eeq
where $y$, $\omega$ and $k_T$ are the rapidity, energy and the
transverse momentum of the radiated quantum in the
nuclear rest frame.
Eq.~(\ref{1.1}) assumes that the radiated energy is relatively
small, i.e. $\omega\ll E_q$. 
It is easy to interpret the formation time (\ref{1.1}) as
lifetime of a photon(gluon)-quark fluctuation \cite{knp}
or as the time needed to distinguish a  radiated quantum from the 
static field of the quark \cite{gw}. 

The total time for bremsstrahlung
is proportional to the initial energy and can therefore substantially 
exceed the time of interaction with the target \cite{hard}.
Radiation continues even after 
the quark leaves the target. This part of radiation 
does not resolve
multiple scattering processes. Important is only
the total momentum transfer. This illuminating manifestation of coherence 
is along these lines that the well known
Landau-Pomeranchuk-Migdal effect (LPM) for long formation times
can be treated. Note that LPM effect corresponds to
the opposite  energy limit, when the radiation time is much shorter 
that the time of propagation through the medium, 
 It was first suggested by Landau and Pomeranchuk \cite{lp}
and investigated by Migdal \cite{m} and has attracted much attention 
during recent years
\cite{n,gw,al,z,McL}. This regime applies only  for the problem
of energy loss in a medium, which is not the problem we discuss here.
Our treatment should apply to the real situation in heavy-ion collisions
at high energies. 
The relationships between the cited papers are complex. In a recent
publication Baier et al. \cite{Bai} have shown that their diagrammatic
approach is in fact equivalent to that of Zakharov \cite{z}. The 
latter is, however, physically far more intuitive and therefore
lends itself more easily to a generalization to the case that
the nuclei are not infinitely extended. 
In another recent paper
Kovchegov and Mueller \cite{Kov} have undertaken the first attempt to
calculate in-medium modification of the transverse momentum
distribution of gluon radiation. This paper  
has also elucidated the relation between
the approaches of \cite{al} and \cite{McL}. In the approach of
\cite{McL} based on the use of the light-cone
gauge the final state interactions summed up in  \cite{al}
(in the covariant gauge) are effectively included in the
light-cone wave function. These observations suggest  that 
all three different approaches might be equivalent when 
followed carefully enough. 

The main goal of this paper is to study the dependence of 
the effects of coherence on
the transverse momentum of the radiated photon or gluon. 
We use the light-cone approach for radiation
first suggested in \cite{hir} and developed in \cite{bhq,z}.
As it is based on an explicit treatment of the transverse 
coordinates it is easily adapted to our purpose. In addition
it seems to be by far the most direct and elegant approach.
We described this approach
in Section~2 for both photon and gluon
bremsstrahlung. We establish a relation between 
the strength of the coherence effects  and the transverse size of 
the Fock state containing the radiated quantum.

The second main result of our paper is the extension of the
light-cone approach to calculations for differential cross sections 
as functions of  
the transverse photon/gluon momentum $\vec k_T$.
This is presented in section~3. 
As one might have  expected, nuclear shadowing,
{\it i.e. suppression} of radiation, is most pronounced at small $k_T$.
An unexpected result is antishadowing, {\it i.e. 
enhancement} of radiation for $k_T > 1\,$GeV, which, however, vanishes 
for still larger  $k_T$.

The results and practical implications for the Drell-Yan process,
prompt photon production and hadroproduction 
are discussed in the last section.

\section{Integrated radiation spectra}

We start with electromagnetic  radiation. 
We cover both, virtual photon radiation (dilepton 
production) and real photon radiation (so called 
prompt photons).

\begin{figure}[tbh]
\includegraphics{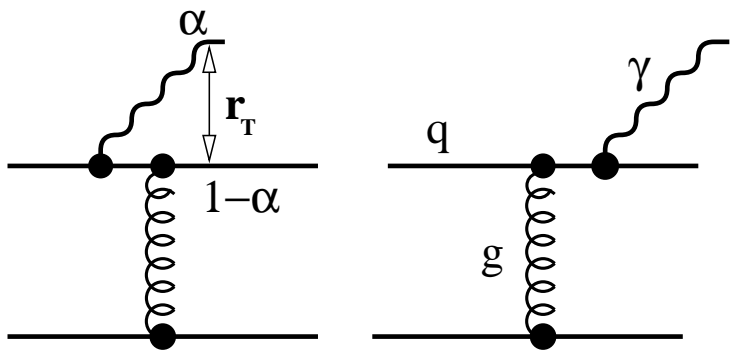}
\begin{center}
\vspace{5cm}
\parbox{13cm}
{\caption[Delta]
{\sl Feynman graphs for bremsstrahlung.}
\label{fig1}}
\end{center}
\end{figure}

The total radiation cross section for (virtual) photons,
as calculated from the 
diagrams shown in Fig.~\ref{fig1}, has the following 
factorized form in impact parameter representation \cite{hir}
(see also \cite{bhq}),
\beq
\frac{d\sigma^N(q\to \gamma q)}{d({\rm ln}\alpha)}=
\int d^2r_T\,\left|\Psi_{\gamma q}(\alpha,\vec r_T)\right|^2\,
\sigma_{\bar qq}(\alpha r_T)\ .
\label{2.1}
\eeq
Here $\Psi_{\gamma q}(\alpha,\vec r_T)$ is the wave
function of the $\gamma-q$ fluctuation of the projectile
quark which depends on $\alpha$, the relative fraction of
the quark momentum carried by the photon, and $r_T$,
the transverse separation between $\gamma$ and $q$
($\Psi$ is  not normalized).
$\sigma_{\bar qq}(\rho)$ is the total interaction cross section for
a $\bar qq$ pair with transverse
separation $\rho$ and a nucleon. $\sigma_{\bar qq}(\rho)$
depends also parametrically on the total collision energy
squared $s$, a dependence we do not write out explicitly
(see, however, section~4).
This becomes only important when fits to actual data are performed.
Eq.(2) contains  a remarkable observation which is 
crucial for this whole approach \cite{hir}:
although we regard only a single projectile quark, the elastic 
amplitude of which is divergent, the {\it radiation} cross 
section is equal
to the total cross section of a $\bar qq$ pair, which is finite.

This can be interpreted as follows.
One should discriminate between the total
interaction cross section and the freeing (radiation)
cross section of a fluctuation. The projectile quark
is represented in the light-cone approach as a sum
of different Fock components. If each of
them interacts with the target with the same
amplitude the coherence between the components is not disturbed,
{\it i.e.} no bremsstrahlung is generated.
Therefore, the production amplitude of a new state (a new
combination of the Fock components) is proportional to
the difference between the elastic amplitudes of different fluctuations.
Thus the universal divergent part of the elastic amplitudes 
cancels and the radiation amplitude is finite.

It is also easy to understand why the $\bar qq$ separation in
(\ref{2.1}) is $\alpha r_T$. As is pointed out 
above one should take the difference
between the amplitudes for a quark-photon fluctuation and a single quark.
The impact parameters of these quarks are different. Indeed,
the impact parameter of the projectile quark serves as the center of gravity 
for the $\gamma-q$ fluctuation in the transverse plane. The
distance to the quark in the quark-gluon Fock-state  
is  then $\alpha r_T$ and that to the photon is 
$(1-\alpha)r_T$.

The wave function of the $\gamma^*q$ fluctuation 
in (\ref{2.1}) for transversely and longitudinally
polarized photons reads (compare with \cite{bk}),
\beq
\Psi^{T,L}_{\gamma^*q}(\vec r_T,\alpha)=
\frac{\sqrt{\alpha_{em}}}{2\,\pi}\,
\chi_f\,\widehat O^{T,L}\,\chi_i\,K_0(\epsilon r_T)
\label{2.2}
\eeq
Here $\chi_{i,f}$ are the spinors of the initial and final quarks.
$K_0(x)$ is the modified Bessel function.
The operators $\widehat O^{T,L}$ have the form,
\beq
\widehat O^{T} = i\,m_q\alpha^2\,
\vec {e^*}\cdot (\vec n\times\vec\sigma)\,
 + \alpha\,\vec {e^*}\cdot (\vec\sigma\times\vec\nabla)
-i(2-\alpha)\,\vec {e^*}\cdot \vec\nabla\ ,
\label{2.2a}
\eeq
\beq
\widehat O^{L}= 2m_{\gamma^*}(1-\alpha)\ ,
\label{2.3}
\eeq
where
\beq
\epsilon^2 = \alpha^2m_q^2 + 
(1-\alpha)m_{\gamma^*}^2\ .
\label{2.3a}
\eeq
$\vec e$ is the polarization vector of the photon,
$\vec n$ is a unit vector along the projectile momentum,
and $\vec\nabla$ acts on $\vec r_T$.
For radiation of prompt photons $m_{\gamma^*}=0$.

Eq.~(\ref{2.1}) can be used for nuclear targets as well.
We consider hereafter formation times given by the energy denominator,
\beq
t_f=\frac{2\,E_q\,\alpha(1-\alpha)}
{\epsilon^2+m_q^2}\gg R_A\ ,
\label{2.3b}
\eeq
which substantially exceed the nuclear radius.
In this limit the transverse $\gamma^*-q$ separation in the 
fluctuation is ''frozen'', {\it i.e.} does not change during 
propagation through the nucleus. The recipe 
for the extension of Eq.~(\ref{2.1}) to a nuclear target 
is quite simple \cite{hir,zkl}. One should just replace 
$\sigma^N_{\bar qq}(\alpha r_T)$ by $\sigma^A_{\bar qq}(\alpha r_T)$,
\beq
\frac{d\sigma^A(q\to \gamma q)}{d({\rm ln}\alpha)}=
2\,\int d^2b
\int d^2r_T\,\left|\Psi_{\gamma q}(\alpha,\vec r_T)\right|^2\,
\left\{1-{\rm exp}\left[
-{1\over 2}\,\sigma_{\bar qq}(\alpha r_T)\,
T(b)\right]\right\}\ ,
\label{2.4}
\eeq
where
\beq
T(b)= \int_{-\infty}^{\infty}
 dz\,\rho_A(b,z)\ .
\label{2.5}
\eeq
Here $\rho_A(b,z)$ is the nuclear density which depends on the 
impact parameter $b$ and the longitudinal coordinate $z$.
One can eikonalize Eq.~(\ref{2.1}) because a fluctuation
with a ''frozen'' transverse size is an eigenstate of interaction 
\cite{zkl}.

Eq.~(\ref{2.4}) shows how  the interference effects work versus $k_T$. 
At small $r_T$ the exponent
$\sigma_{\bar qq}(\alpha r_T)T(b)/2 \ll 1$ since 
$\sigma_{\bar qq}(\alpha r_T)$ is small. Therefore, one
can expand the exponential and the cross section turns out to be 
proportional to $A$. This is the Bethe-Heitler limit for bremsstrahlung.
In the opposite limit $\sigma_{\bar qq}(\alpha r_T)T(b)/2 \gg 1$
one can neglect the exponential for $b\leq R_A$ and the cross section
(\ref{2.4}) is proportional to $A^{2/3}$. This is the limit of full coherence
when the whole row of nucleons with the same impact parameter acts 
like a single nucleon. As the 
gluon transverse momentum is related to the inverse of 
$r_T$,  one could expect that the limit of maximal coherence 
is reached for small $k_T$, and the Bethe-Heitler limit for
large $k_T$. The situation is, however, more complicated
as discussed in the next section.

\medskip

\begin{figure}[tbh]
\includegraphics{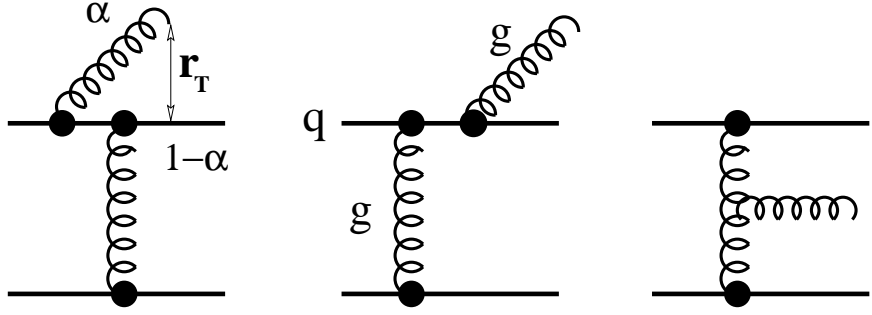}
\begin{center}
\vspace{5cm}
\parbox{13cm}
{\caption[Delta]
{\sl Feynman graphs for gluon bremsstrahlung of an
interacting quark.}
\label{fig2}}
\end{center}
\end{figure}

Gluon radiation is described by the diagrams 
\cite{gb} shown in 
Fig.~\ref{fig2}.
The radiation cross section for a nucleon target
and the nuclear effects \cite{hir}
look similar to those of 
Eqs.~(\ref{2.1}) - (\ref{2.4}) 
\beq
\frac{d\sigma^A(q\to gq)}{d({\rm ln}\alpha)}=
2\,\int d^2b
\int d^2r_T\,\left|\Psi_{g q}(\alpha,\vec r_T)\right|^2\,
\left\{1-{\rm exp}\left[
-{1\over 2}\,\sigma_{g\bar qq}(\vec r_1,\vec r_2)\,
T(b)\right]\right\}\ ,
\label{2.6}
\eeq
where 
$\Psi_{g q}(\alpha,\vec r_T)$ 
is the wave function of a quark-gluon fluctuation which has
the same form as in Eq.(\ref{2.2}), but with the 
replacements $\gamma^*\Rightarrow g$,
$\alpha_{em}\Rightarrow 4\alpha_s/3$ and $m_{\gamma^*}\Rightarrow m_g$.
We keep the gluon mass nonzero in order to simulate the possible effects
of confinement on gluon bremsstrahlung.
$\sigma_{g\bar qq}$
is the interaction cross section of a 
colorless $g\bar qq$ system with a nucleon \cite{nz},
\beq
\sigma_{g\bar qq}(\vec r_1,\vec r_2)=
\frac{9}{8}\Bigl\{\sigma_{\bar qq}(r_1) +
\sigma_{\bar qq}(r_2)\Bigl\}-\frac{1}{8}\,
\sigma_{\bar qq}(\vec r_1-\vec r_2)\ ,
\label{2.8}
\eeq
where $\vec r_1$ and $\vec r_2$ are the transverse separations 
gluon -- quark and gluon -- antiquark respectively.
In the case of gluon radiation, i.e.  Eq.~(\ref{2.6}),  $\vec r_1=\vec
r_T$
and
$\vec r_2=(1-\alpha)\vec r_T$.  

Although Eq.~(\ref{2.6}) looks simple, it includes 
the effects of quark and gluon
rescattering in the nucleus to all orders.

\section{The transverse momentum distribution}

\subsection{Electromagnetic radiation}

The transverse momentum distribution of photon bremsstrahlung in 
quark-nucleon interactions integrated over the final quark transverse momentum
reads (see Appendix A),
\beq
\frac{d^3\sigma^N(q\to q\gamma)}
{d({\rm ln}\alpha)\,d^2k_T} =
\frac{1}{(2\pi)^2}
\int d^2r_1\,d^2r_2\,
{\rm exp}\bigl[i\vec k_T(\vec r_1-\vec r_2)\bigr]\,
\Psi_{\gamma q}^*(\alpha,\vec r_1)\,
\Psi_{\gamma q}(\alpha,\vec r_2)\,
\sigma_{\gamma}(\vec r_1,\vec r_2,\alpha)\ ,
\label{3.1}
\eeq
where
\beq
\sigma_{\gamma}(\vec r_1,\vec r_2,\alpha)={1\over 2}\Bigl\{
\sigma_{\bar qq}(\alpha r_1) + \sigma_{\bar qq}(\alpha r_2) - 
\sigma_{\bar qq}[\alpha (\vec r_1-\vec r_2)]\Bigr\}\ .
\label{3.1a}
\eeq
By integrating over $k_T$ one obviously recovers Eq.~(\ref{2.1}),
since $\sigma_{\gamma}(\vec r,\vec r,\alpha)=\sigma_{\bar qq}(\alpha r)$.

For $\alpha\ll 1$ one can use the dipole approximation
for the cross section, i.e. one can set 
$\sigma_{\bar qq}(\rho) = C\,\rho^2$.
Moreover, this approximation  works also rather well at larger interquark 
separations, even for hadronic sizes \cite{hp}. For the latter the cross
section is proportional to  the mean radius squared. Therefore, we
use the dipole approximation for all cases considered. 
Then (\ref{3.1a}) simplifies to
\beq
\sigma_{\gamma}(\vec r_1,\vec r_2,\alpha)\approx
C\,\alpha^2\,\vec r_1\cdot \vec r_2\ ,
\label{3.1b}
\eeq
and we can explicitly calculate the $k_T$ distribution (\ref{3.1}),
\beqn
\frac{d^3\sigma^N_T(q\to q\gamma^*)}
{d({\rm ln}\alpha)\,d^2k_T} &=&
\frac{\alpha_{em}}{\pi^2}\,
\frac{C\,\alpha^2}{(k_T^2+\epsilon^2)^4}\,
\Bigl\{2\,m_q^2\,\alpha^4\,k_T^2 +
\bigl[1+(1-\alpha)^2\bigr](k_T^4+\epsilon^4)\Bigr\}\ ,
\label{3.1c}\\
\frac{d^3\sigma^N_L(q\to q\gamma^*)}
{d({\rm ln}\alpha)\,d^2k_T} &=& 
\frac{4\,\alpha_{em}\,C\,\alpha^2(1-\alpha)^2\,
m^2_{\gamma^*}\,k_T^2}
{\pi^2\, (k_T^2+\epsilon^2)^4}\ .
\label{3.1d}
\eeqn

Note that for small $\alpha$ (\ref{3.1c}) and (\ref{3.1d}) vanish 
like  $\alpha^2$. This could have been expected
since electomagnetic bremsstrahlung is known
to be located predominantly in the fragmentation regions of 
colliding particles 
rather than at midrapidity.

In the case of a nuclear target the transverse momentum
distribution has to be modified by eikonalization of (\ref{3.1}) 
(see Appendix A),
\beq
\frac{d^3\sigma^A(q\to q\gamma)}
{d({\rm ln}\alpha)\,d^2k_T} =
\frac{1}{(2\pi)^2}
\int d^2r_1\,d^2r_2\,
{\rm exp}\bigl[i\vec k_T(\vec r_1-\vec r_2)\bigr]\,
\Psi_{\gamma q}^*(\alpha,\vec r_1)\,
\Psi_{\gamma q}(\alpha,\vec r_2)\,
\Sigma_{\gamma}(\vec r_1,\vec r_2,\alpha)\ ,
\label{3.2}\\
\eeq
where
\beqn
\Sigma_{\gamma}(\vec r_1,\vec r_2,\alpha)&=&
\int d^2b\biggl\{1+
{\rm exp}\Bigl[-{1\over 2}\,
\sigma_{\bar qq}[\alpha (\vec r_1-\vec r_2)]\,
T(b)\Bigr]\nonumber\\
&-& {\rm exp}\Bigl[-{1\over 2}\,
\sigma_{\bar qq}(\alpha r_1)\,T(b)\Bigr]
 - {\rm exp}\Bigl[-{1\over 2}\,
\sigma_{\bar qq}(\alpha r_2)\,T(b)\Bigr]\biggr\}
\label{3.3}
\eeqn
The fluctuation wave functions in (\ref{3.2})  can be represented using
(\ref{2.2}) in the form
\beqn
\sum\limits_{in,\,f}\Psi^T_{\gamma^* q}(\vec r_1,\alpha)\,
{\Psi^T}^*_{\gamma^* q}(\vec r_2,\alpha) &=&
\frac{\alpha_{em}}{2\,\pi^2}\,\Bigl\{m_q^2\,\alpha^4\,
K_0(\epsilon r_1)\,K_0(\epsilon r_2) \nonumber\\
&+& \left[1+(1-\alpha)^2\right]\,\epsilon^2\,
\frac{\vec r_1\vec r_2}{r_1\,r_2}\,
K_1(\epsilon r_1)\,K_1(\epsilon r_2)\Bigl\}\ ,
\label{3.3a}
\eeqn
\beq
\sum\limits_{in,\,f}\Psi^L_{\gamma^* q}(\vec r_1,\alpha)\,
{\Psi^L}^*_{\gamma^* q}(\vec r_2,\alpha) = 
\frac{2\,\alpha_{em}}{\pi^2}\,
m_{\gamma^*}^2\,(1-\alpha)^2\,
K_0(\epsilon r_1)\,K_0(\epsilon r_2)\ ,
\label{3.3b}
\eeq
where we average over the initial quark polarization and sum over 
the final polarizations of quark and photon.

At first glance, one could think that 
the $k_T$ distribution is not modified 
by the nucleus in the case $t_f\gg R_A$, since the fluctuation
is formed long before the nucleus 
and the quark interact.
This is, however, not the case. Due to
color filtering \cite{bbgg} the mean size of $\bar qq$ dipoles
surviving propagation through the nucleus decreases with $A$.
Correspondingly, the transverse momentum of the photon increases.
In other words, a heavier nucleus provides a larger momentum transfer 
to the quark, hence it is able to break up smaller size fluctuations
and release photons with larger $k_T$.

Note that one can also calculate the distribution 
with respect to the  transverse momentum
$\vec p_T$ of the final quark integrating 
the differential cross section over 
the photon momentum $\vec k_T$. The result turns out to be the same as
(\ref{3.1}) and (\ref{3.2}) with the replacement $\alpha \Rightarrow 
1-\alpha$. 

We also calculated the nuclear dependence of the differential cross section
(\ref{3.2}) - (\ref{3.3}) using the dipole approximation for 
$\sigma_{\bar qq}(r)$. The details of the necessary 
integration can be found in 
Appendix~B. As usual, we approximate the cross section
by an $A^n$-dependence. The power $n$ is then defined by
\beq
n(k_T,\alpha) = \frac{d\left\{{\rm ln}
\left[d^3\sigma^A(q\to q\gamma)/
\Bigl(d({\rm ln}\alpha)\,d^2k_T\Bigr)\right]\right\}}
{d({\rm ln}A)}
\label{3.4}
\eeq
This power can also be $A$ dependent. We performed calculations for
$A= 200$. To simplify these calculations, we
used the constant density distribution, $\rho_A(r)=\rho_0
\Theta(R_A-r)$ with $\rho_0=0.16\,$fm$^{-3}$.

\begin{figure}[tbh]
\includegraphics{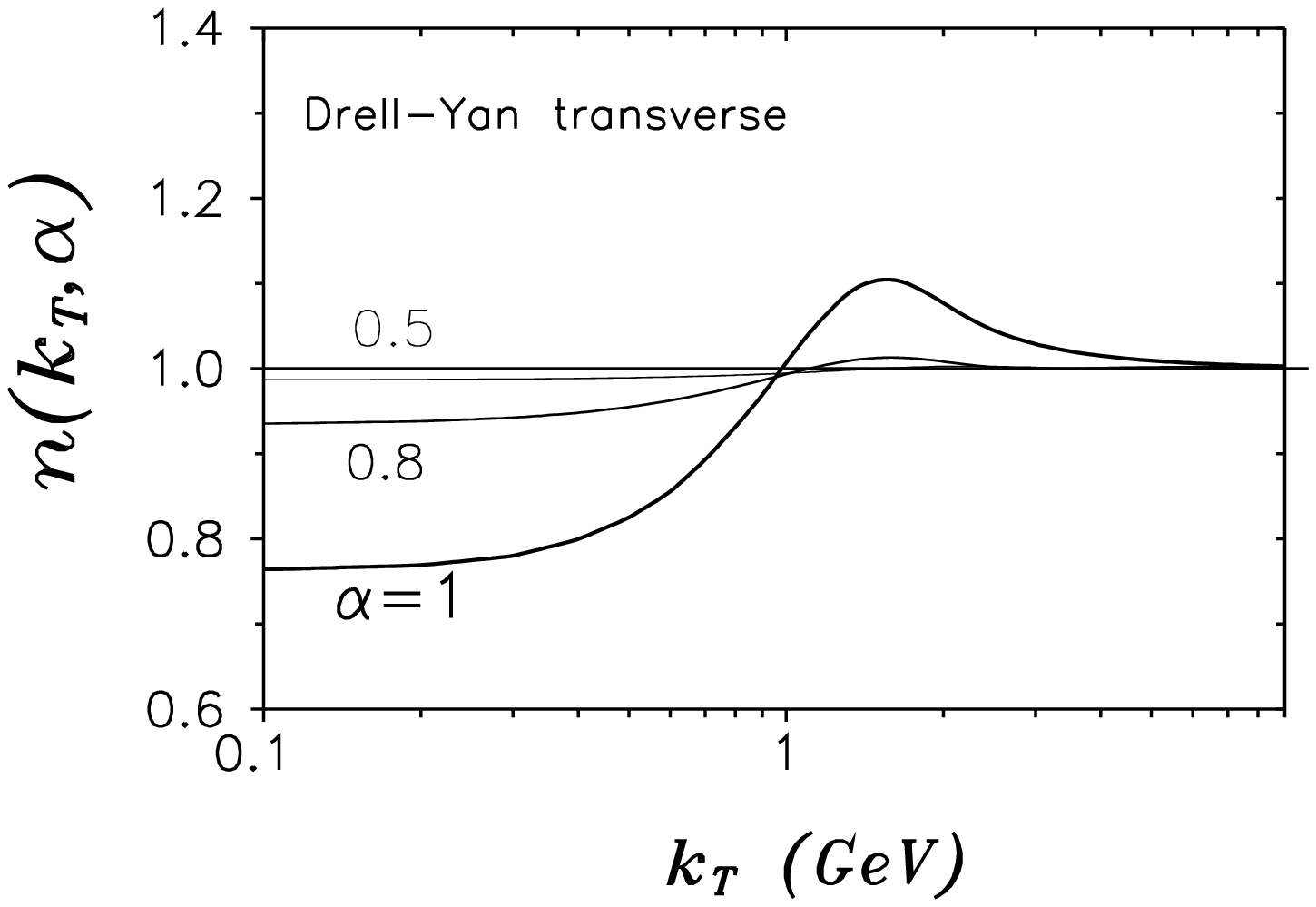}
\includegraphics{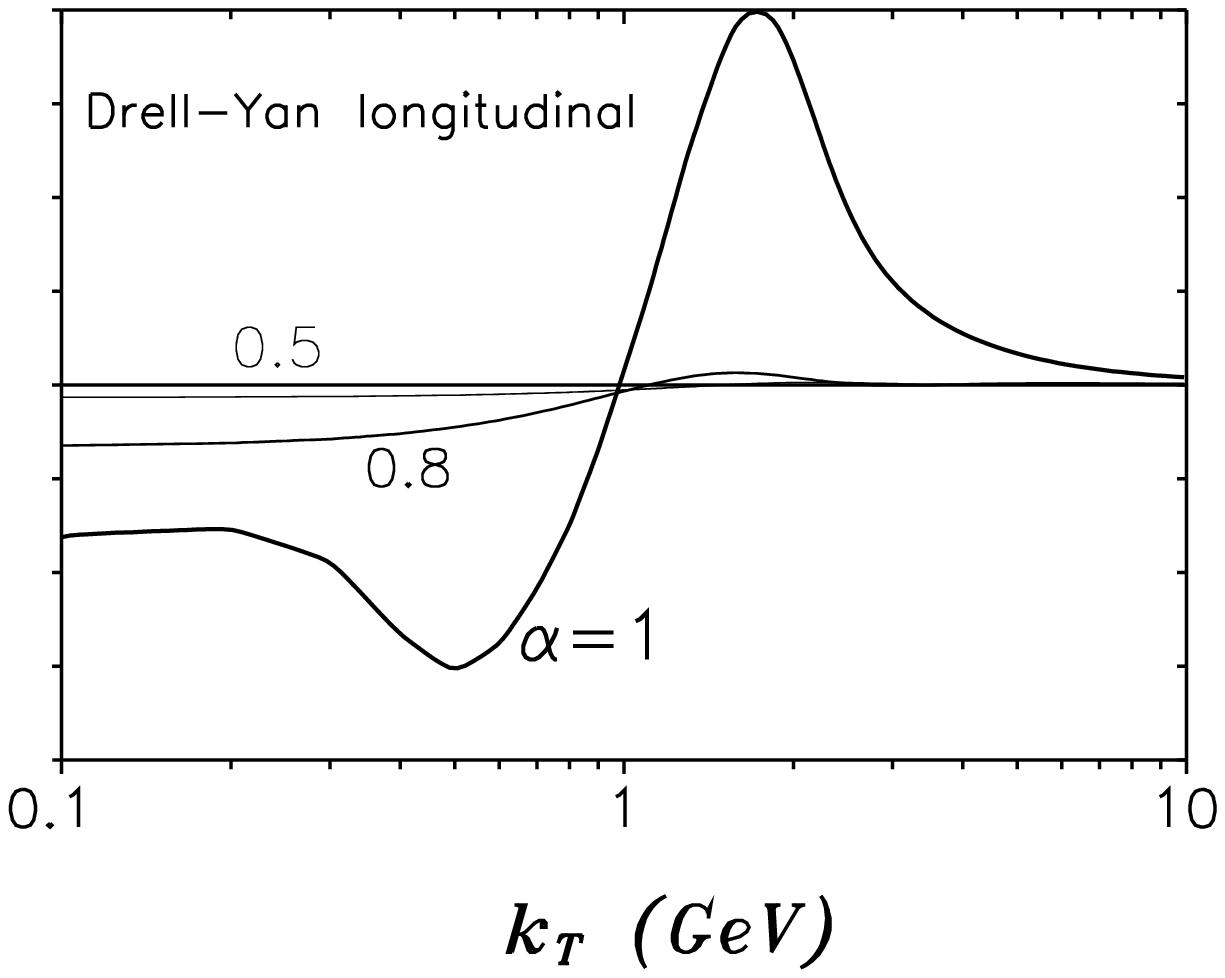}
\begin{center}
\vspace{6.5cm}
\parbox{13cm}
{\caption[Delta]
{\sl The exponent (\ref{3.4}) of the atomic
number dependence parameterized as $A^n$ versus
$k_T$ and $\alpha$ for transversely (left figure) and longitudinally
(right figure) polarized virtual photons.}
\label{fig3}}
\end{center}
\end{figure}

First of all, we calculated $n(k_T,\alpha)$ for Drell-Yan
lepton pair production at $m_{\gamma^*}=4\,$GeV. 
The results are shown in Fig.~\ref{fig3} for transversely 
and longitudinally polarized virtual photons (the two 
components can be extracted from the angular distribution
of the lepton pairs).
We see that $n < 1$ for $k_T < 1\,$GeV, {{\it i.e.}
the Drell-Yan pair production is shadowed by the nucleus.
The shadowing is stronger for larger $\alpha$ \cite{hir}. 
Shadowing in the Drell-Yan 
process was first observed  by the E772 Collaboration \cite{e772}.
Their effect is, however, much weaker which can easily be explained
because  for Fermilab energies the 
radiation time (\ref{1.1}) is quite short compared to the nuclear radius.
This fact is taken into account in \cite{hir} by means of nuclear 
formfactor. Then the data can be described quite nicely.
(See also \cite{hk-dy}.)

An interesting result contained in Fig.~\ref{fig3} is 
the appearance of an antishadowing region for $k_T > 1\,$GeV. 
This is the first case in which the coherence effects 
enhances rather than suppresses the
radiation spectrum.
It originates from an interference effect which is not noticeable for
the integrated quantities. 

Nuclear antishadowing is especially strong for longitudinal 
photons and $k_T \sim 1.5 - 2\,$ GeV. Color filtering
in nuclear matter changes the angular distribution
of Drell-Yan pairs and enhances the yield of longitudinally polarized dileptons.
The nontrivial behaviour of $n$ for longitudinal photons at
small $k_T$ is due to the dip at $k_T=0$ in the differential cross section 
for a nucleon, see Eq. (\ref{3.1d}) . This minimum is filled by multiple
scattering of the quark in the nucleus leading to an increase of
$n(k_T=0)$ and a strong $A$-dependence of $n(k_T=0)$.
(Formally, for longitudinal photons  $n(k_T=0)$ goes to infinity for
$A=1$, because the proton cross-section at $k_T=0$ is zero). 

Note that nuclear enhancement of Drell-Yan pair production at large $k_T$ 
was also observed experimentally \cite{e772}. However, as was mentioned,
these data were taken  in
the kinematical region of the Bethe-Heitler regime, i.e.  $t_f \ll R_A$.
Therefore, they cannot be compared with our calculations. In fact the
observation was explained quite satisfactory 
in \cite{hk-dy}.

The $k_T$-dependence of $n$ is expected to be nearly the same for different
dilepton masses, down to the mass range probed in  the 
CERES experiment at SPS CERN. However, the nuclear effects turn out
to be
quite different for real photons. Our results are shown in 
Fig.~\ref{fig4}.
\begin{figure}[tbh]
\includegraphics{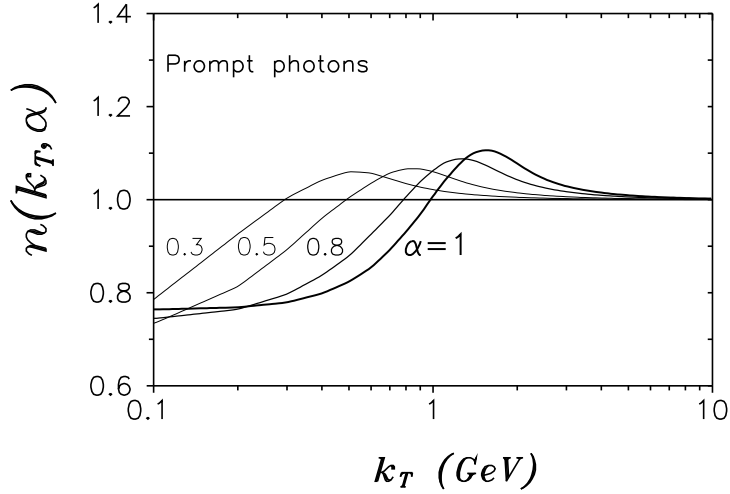}
\begin{center}
\vspace{6.5cm}
\parbox{13cm}
{\caption[Delta]
{\sl The same as in Fig.~\ref{fig3}, but for real photons.}
\label{fig4}}
\end{center}
\end{figure}
In order to compare with experimental dilepton
cross sections and prompt photon
production rates our results have to be convoluted 
with the quark distribution function for
the projectile proton. Since the electromagnetic radiation steeply
falls off with decreasing $\alpha$
(proportional to $\alpha^2$, see (\ref{3.1c}) - (\ref{3.1d})), 
the convolution effectively picks out large
values of $\alpha$ where the nuclear effects are in turn expected to be large.
Detailed calculations and comparisons with data are postponed to a
later publication. 

\subsection{Gluon radiation}

Now we can discuss bremsstrahlung in the non-Abelian case.
Summing up the diagrams in Fig.~\ref{fig2} we get in
impact parameter representation 
\beq
\frac{d^3\sigma^N(q\to qg)}
{d({\rm ln}\alpha)\,d^2k_T} =
\frac{1}{(2\pi)^2}
\int d^2r_1\,d^2r_2\,
{\rm exp}\bigl[i\vec k_T(\vec r_1-\vec r_2)\bigr]\,
\Psi_{gq}^*(\alpha,\vec r_1)\,
\Psi_{gq}(\alpha,\vec r_2)\,
\sigma_g(\vec r_1,\vec r_2,\alpha)\ ,
\label{4.1}
\eeq
where (see Appendix A)
\beq
\sigma_g(\vec r_1,\vec r_2,\alpha)={1\over 2}
\Bigl\{\sigma_{g\bar qq}
(\vec r_1,\vec r_1-\alpha r_2) + 
\sigma_{g\bar qq}(\vec r_2,\vec r_2-\alpha r_1) 
-\sigma_{\bar qq}[\alpha (\vec r_1-\vec r_2)] -
\sigma_{gg}(\vec r_1-\vec r_2)
\Bigr\}\ .
\label{4.2}
\eeq
Here $\sigma_{gg}(r)={9\over 4}\,\sigma_{\bar qq}(r)$ is the 
total cross section
of a colorless $gg$ dipole with a nucleon.

Note that  (\ref{4.2}) reproduces several simple limiting cases:\\
1.) $\sigma_g(\vec r_1,\vec r_2,\alpha)$
vanishes if either of $r_1$ or $r_2$ goes to zero, which
expresses the fact that a point-like quark-gluon fluctuation cannot 
be resolved by any interaction. To show this limiting behaviour one
simply has to insert e.g. for $\vec r_2=0$ 
the two relations $\sigma_{g\bar q q}(\vec r_1,
\vec r_1)= \sigma_{gg}(\vec r_1)$ and  $\sigma_{g\bar q q}(\vec 0,
-\alpha\vec r_1)= \sigma_{\bar qq}(-\alpha\vec r_1)=
\sigma_{\bar qq}(\alpha\vec r_1)$. (Quark and antiquark at the same 
point in space act like a gluon etc.)\\
2.) For $\alpha\to 1$ the quark-gluon separation tends to zero
and (\ref{4.2}) transforms into (\ref{3.1a}). On the
other hand, at $\alpha\to 0$ the quark-antiquark separation
vanishes and (\ref{4.2}) takes again the same form as (\ref{3.1a}),
except that the $\bar qq$ pair is replaced by a gluon-gluon dipole.
\beq
\sigma_g(\vec r_1,\vec r_2,\alpha)\bigr|_{\alpha\ll 1}=
{1\over 2}\Bigl\{\sigma_{gg}(r_1) + \sigma_{gg}(r_2) -
\sigma_{gg}[(\vec r_1-\vec r_2)]\Bigr\}
=\frac{9}{4}\,\sigma_{\gamma^*}
(\vec r_1,\vec r_2,\alpha)\Bigr|_{\alpha=1}\ .
\label{4.3}
\eeq

We use the dipole approximation $\sigma_{\bar qq}(r_T)
\approx C\,r_T^2$, which is well justified in this case
since the mean transverse quark-gluon separation
is small at small $\alpha$.
In this case (\ref{4.2}) and (\ref{2.8}) lead to
\beq
\sigma_g(\vec r_1,\vec r_2,\alpha)\approx
\left[\alpha^2+{9\over 4}(1-\alpha)\right]\,C\,
\vec r_1 \cdot \vec r_2
\label{4.4}
\eeq
This expression coincides with (\ref{3.1b}) 
up to the factor $[1+9(1-\alpha)/(4\alpha^2)]$.
Therefore, we can use the results (\ref{3.1c}) - 
(\ref{3.1d}) obtained for 
photon bremsstrahlung which 
for $\alpha\to 0$ lead to
\beq
\frac{d^3\sigma^N_T(q\to qg)}
{d({\rm ln}\alpha)\,d^2k_T}\biggr|_{\alpha\ll 1}
\approx
\frac{6\,C\,\alpha_{s}}{\pi^2}\ 
\frac{k_T^4+m^4_{g}}
{(k_T^2+m^2_{g})^4}
\label{4.5}
\eeq
\beq
\frac{d^3\sigma^N_L(q\to qg)}
{d({\rm ln}\alpha)\,d^2k_T}\biggr|_{\alpha\ll 1}
\approx \frac{12\,C\,\alpha_{s}\,
m^2_{g}\,k_T^2}
{\pi^2\,(k_T^2+m_g^2)^4}
\label{4.6}
\eeq
In contrast to  photon bremsstrahlung this cross 
sections do not vanish for $\alpha\to 0$. 
This is a consequence of the non-Abelian 
nature of QCD \cite{gb}. 
The radiating color current propagates through the whole rapidity
interval between the projectile and the target providing
a constant gluon density (\ref{4.5}) - (\ref{4.6}) with respect to
rapidity.

Eikonalization of the cross section (\ref{4.1}) results in,
\beq
\frac{d^3\sigma^A(q\to qg)}
{d({\rm ln}\alpha)\,d^2k_T} =
\frac{1}{(2\pi)^2}
\int d^2r_1\,d^2r_2\,
{\rm exp}\bigl[i\vec k_T(\vec r_1-\vec r_2)\bigr]\,
\Psi_{gq}^*(\alpha,\vec r_1)\,
\Psi_{gq}(\alpha,\vec r_2)\,
\Sigma_g(\vec r_1,\vec r_2,\alpha)\ ,
\label{4.7}
\eeq
where
\beqn
&&\Sigma_g(\vec r_1,\vec r_2,\alpha)=
\int d^2b\biggl\{{\rm exp}\Bigl[-{1\over 2}
\sigma_{\bar qq}[\alpha (\vec r_1-\vec r_2)]\Bigr] +
{\rm exp}\Bigl[-{1\over 2}
\sigma_{gg}(\vec r_1-\vec r_2)\,T(b)\Bigr]
\nonumber\\&-&{\rm exp}\Bigl[-{1\over 2}
\sigma_{g\bar qq}(\vec r_1,\vec r_1-\alpha r_2)\,T(b)\Bigr] -
{\rm exp}\Bigl[-{1\over 2}
\sigma_{g\bar qq}(\vec r_2,\vec r_2-\alpha r_1)\,T(b)\Bigr] 
\biggr\}
\label{4.8}
\eeqn

In the limit $\alpha\ll 1$, which is
of practical interest at high energy 
(\ref{4.2}) transforms to the form of (\ref{4.3}) 
and Eq.~(\ref{4.8}) simplifies to
\beqn
\Sigma_g(\vec r_1,\vec r_2,\alpha)\bigr|_{\alpha\ll 1}&=&
\int d^2b\biggl\{1 +
{\rm exp}\Bigl[-{1\over 2}
\sigma_{gg}(\vec r_1-\vec r_2)\,T(b)\Bigr]
\nonumber\\
& - &{\rm exp}\Bigl[-{1\over 2}
\sigma_{gg}(\vec r_1)\,T(b)\Bigr] -
{\rm exp}\Bigl[-{1\over 2}
\sigma_{gg}(\vec r_2)\,T(b)\Bigr]
\biggr\}
\label{4.9}
\eeqn
Note that the transverse momentum distribution for gluon radiation
was calculated previously in \cite{Kov} in the limit $\alpha \to 0$
and $m_q=m_g=0$. Our results (\ref{4.7}), (\ref{4.9}) agree with
that in \cite{Kov} in this limit.

In (\ref{4.9}) we make use of the fact that at
zero $\bar qq$ separation
a $g\bar qq$-system interacts like a pair of gluons,
$\sigma_{g\bar qq}(\vec r,\vec r) = \sigma_{gg}(r) =
(9/4) \sigma_{\bar qq}(r)$. Therefore, (\ref{4.7}) - (\ref{4.8}) can be
calculated in the same way as (\ref{3.1}) - (\ref{3.2}) in the electromagnetic
case at $\alpha=1$ (see Appendix~B), except that the fluctuation wave functions
must be taken at $\alpha=0$. We assign an effective mass to the gluon,
either of the order of the inverse confinement radius, $m_g\approx 0.15\,$GeV,
or in accordance with
the results of lattice calculations for the range of gluon-gluon
correlations \cite{sh} of size  $m_g=0.75\,$GeV. We sum over the polarization
of the
emitted gluon.
The numerical results are
plotted in Fig.~\ref{fig5}.
\begin{figure}[tbh]
\includegraphics{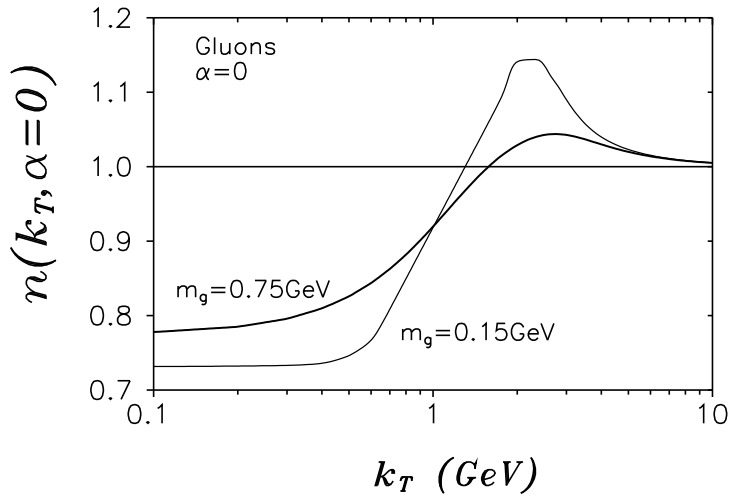}
\begin{center}
\vspace{6.5cm}
\parbox{13cm}
{\caption[Delta]
{\sl The same as in Fig.~\ref{fig3}, but for gluons
at $\alpha=0$ for different effective gluon masses.}
\label{fig5}}
\end{center}
\end{figure}
They are qualitatively similar to those 
for photon radiation (see Fig.~\ref{fig3}): 
shadowing at small and antishadowing at large 
$k_T$. However, the effect of antishadowing is more pronounced for 
light gluons.

Antishadowing of gluons results in antishadowing for inclusive
hadron production, which is well known as Cronin effect \cite{cronin1}.
Although it was qualitatively understood that the source of 
this enhancement
is multiple interaction of the partons in the nucleus, 
to our knowledge no 
realistic calculation taking into account color screening was
done so far. We expect that the 
Cronin effect disappears at very large $k_T$,
which would actually be in accordance with 
available data \cite{cronin2}.
For a honest  comparison with these data, one  
has to  relate the  $k_T$ of the gluon to that
of the produced hadron, a step which lies not within the scope of
this paper.

\section{Conclusions and discussion}
 
The main results of the paper are the following.
\begin{itemize}
\item
The factorized light-cone approach \cite{hir} for 
the analysis of radiation cross sections was
extended to treat the  $k_T$ dependence, and
was applied 
both to photon (real and virtual) and gluon bremsstrahlung.

\item
The effects of coherence which are known to 
suppress radiation at long formation times, 
is only effective for  small $k_T$. At $k_T>1\,$GeV the interference 
instead actually enhances the
radiation spectrum. This was indeed observed for dilepton
and inclusive hadron production off nuclei (Cronin effect).
The enhancement of radiation by the coherence effects 
turns out to vanish at very large
transverse momenta $k_T\geq 10\,$GeV. This was also
observed in hadroproduction .

\item
suppression and enhancement of radiation by the effects of coherence 
are quite different for 
transversely and longitudinally polarized photons.
Both contributions can be separated by 
measuring the  angular distribution of the produced dileptons.

\end{itemize}

Note that we use Born graphs shown in Figs.~\ref{fig1} - \ref{fig2}
to derive expressions (\ref{2.1}) and others having a factorized form.
As a result of Born approximation the dipole cross section 
$\sigma_{\bar qq}(\rho)$ is energy independent. It is well known \cite{bfkl}
that the higher order corrections lead to a cross section rising with energy. 
HERA data suggest that this energy
dependence is correlated with the dipole size $r_T$. Therefore,
the parameter $C(s)$ can be parameterized as  
\beq
C(s)=C_0\,\left(\frac{s}{s_0}\right)^{\Delta(r_T)}\ ,
\label{5.1}
\eeq
where $s_0=100\,$GeV$^2$, $C_0\approx 3$.
The power $\Delta(r_T)$ grows with decreasing $r_T$. 
This dependence is extracted from an 
analysis of HERA data in \cite{kp98}

Our results obtained for the radiation by a quark interacting
with a nucleus are easily adapted to proton--nucleus collisions
by convolution with the quark distribution in the proton.

We plan also to extend our analysis to relativistic heavy ion collisions.
The condition we use, $t_f\gg R_A$ is poorly satisfied at
present fixed target accelerators, but are well justified at
RHIC or LHC. Indeed, 
if $s_{NN}$ is the total $NN$ collision energy squared, for a gluon(photon)
radiated at central rapidity,
\beq
\alpha = \frac{3\,k_T}{\sqrt{s_{NN}}}
\label{5.2}
\eeq
\beq
t_f=\frac{\sqrt{s_{NN}}}{m_N\,k_T}\ .
\label{5.3}
\eeq
We conclude that at RHIC or LHC energies $\alpha\ll 1$
and that gluons with a few GeV transverse momentum
are radiated far away from the nucleus, i.e. $t_f\gg R_A$.
Thus our calculations should be directly applicable.

{\bf Acknowledgements:} We are grateful to J\"org
H\"ufner for many stimulating
and fruitful discussions and to Vitali Dodonov for help
with numerical calculations.
We are especially thankful to Urs Wiedemann whose questions 
helped us to make the presentation more understandable. He also
found a few misprints in Appendix A.
The work of A.V.T was supported by the Gesellschaft f\"ur
Schwerionenforschung, GSI, grant HD H\"UF T,
and A.S. was supported by the GSI grant OR SCH T.
A.V.T. and A.S.  greatly acknowledge the hospitality of the MPI
f\"ur Kernphysik.

\def\appendix{\par
 \setcounter{section}{0}
 \setcounter{subsection}{0}
 \def\thesection{Appendix \Alph{section}}
 \def\thesubsection{\Alph{section}.\arabic{subsection}}
 \def\theequation{\Alph{section}.\arabic{equation}}
 \setcounter{equation}{0}}

\appendix

\section{}

\setcounter{equation}{0}

In this section we illustrate how to eikonalize the differential
 cross
section in the case of a  nuclear target and for the example of
 electomagnetic
bremsstrahlung of an electron. The latter is described as  propagating in a
 stationary field $U(\vec x)$,
where $\vec x$ is a three-dimensional
 vector.

 The differential cross
section reads,
 \beq
 \frac{d^5\sigma}{d({\rm ln}\alpha)\,d^2p_T\,d^2k_T} =
\frac{\alpha_{em}}{(2\pi)^4}\,
 \left|M_{fi}\right|^2\ ,
\label{a.1}
\eeq
 where $\vec k_T$ and $\vec p_T$ are the transverse momenta
 of the
photon and the electron in the final state.

 The radiation amplitude for
a transversely polarized massive
 photon $\gamma^*$
($\omega^2=k^2+m_{\gamma^*}^2$) has the form,
 \beq
 M_{fi}^T = \int
d^3x\,{\Psi^-}^{\dagger}(\vec x,\vec p_2)\,
 ~\widehat{\vec \alpha}\cdot \vec e^*\ 
e^{-i\vec k\vec x}\,
 \Psi^+(\vec x,\vec p_1)\ ,
\label{a.2}
\eeq
 where $\widehat{\vec \alpha}=\gamma_0\vec\gamma$ are the Dirac
matrices, and
 the wave functions $\Psi(\vec x,\vec p_{1,2})$ of the initial
and final electron, are solutions of the Dirac equation in
 the
external potential $U(\vec x)$,
 \beq
 \left[\epsilon_{1,2} - U(\vec x) -
m\,\beta +
 i\,\widehat{\vec \alpha}\vec\nabla\right]\,
 \Psi(\vec x,\vec
p_{1,2}) = 0\ .
\label{a.3}
\eeq
 The upper indices ''$ - $'' and ''$+$'' in (\ref{a.2}) indicate that for the
initial and final states the solutions contain in addition
to the plane wave also an outgoing and incoming spherical wave respectively.

If the energy is sufficiently high, $\epsilon_{1,2}\gg m,\,U$ it is natural 
to search for a solution of (\ref{a.3}) in the form of a polynomial expansion 
over powers of $1/\epsilon$ ($\epsilon=\epsilon_{1,2}$),
\beqn
&&\Psi(\vec x,\vec p) = \sum\limits_{n-0}^{\infty}\Psi_n(\vec x,\vec p)\ ,
\nonumber\\
&&\Psi_n(\vec x,\vec p) \sim \epsilon^{-n}\ .
\label{a.3a}
\eeqn

Note that in the case of radiation of a longitudinally polarized 
photon it is sufficient to take into account only the main ($\Psi_0$)
which has a form,
\beq
\Psi_0(\vec x,\vec p) = e^{i\vec p\vec r}\,
f(\vec x,\vec p)\,\frac{u(\vec p)}{\sqrt{2\epsilon}}\ ,
\label{a.3b}
\eeq
where $u(\vec p_{1,2})$ is the 4-component spinor corresponding to
a free electron with momentum $\vec p_{1,2}$,
and the scalar function $f(\vec x,\vec p)$ is a solution of the equation,
\beq
\left(\Delta + 2\,i\,\vec p\,\vec\nabla -
2\,\epsilon\,U(\vec x)\right)\,f(\vec x,\vec p) = 0\ .
\label{a.3c}
\eeq

In the case of radiation of transversely polarized photons it is
known \cite{ll} that the two first terms in expansion (\ref{a.3a})
are important. Their sum can be represented in the form,
of Furry approximation, $\Psi_F$ \cite{furry}
\beq
\Psi_0 + \Psi_1 \equiv \Psi_F=
e^{i\vec p\vec x}\,\left(1 -
\frac{i\widehat{\vec\alpha}\vec\nabla}{2\epsilon}\right)\,
f(\vec x,\vec p)\,\frac{u(\vec p)}{\sqrt{2\epsilon}}\ .
\label{a.3d}
\eeq

One can estimate the accuracy of the Furry approximation using 
the following relations,
\beq
\Psi - \Psi_F \equiv \delta\Psi= 
e^{i\vec p\vec x}\, \Phi(\vec x,\vec p)\ ,
\label{a.3e}
\eeq
where $\Phi(\vec x,\vec p)$ satisfies the equation,
\beq
\Bigl[\Delta+2i\vec p\cdot\vec\nabla -
2\epsilon\,U(\vec x) + i\vec\alpha\cdot\vec\nabla\,
U(\vec x)\Bigr]\,\Phi(\vec x,\vec p) = 
- \frac{1}{2\epsilon}\,
\Bigl(\vec\alpha\cdot\vec\nabla\,U(\vec x)\Bigr)
\Bigl(\vec\alpha\cdot\vec\nabla\,f(\vec x,\vec p)\Bigr)
\label{a.3f}
\eeq

It turns out that this correction to the Furry approximation
for the electron wave function is of the order of $\bar U/\epsilon$
in the bremsstrahlung cross section.

It is convenient (see below) to chose the axis $z$ along the momentum of the
radiated photon. 
In this case one can represent the Furry approximation (\ref{a.3d})
for the functions $\Psi^+(\vec x,\vec p_1)$ and $\Psi^-(\vec x,\vec p_2)$
in the form,
 \beq
 \Psi^+_F(\vec x,\vec p_1) =
 e^{ip_1z}\,
 \hat D_1\,F^+(\vec x,\vec
p_1)\,
 \frac{u(\vec p_1)}{\sqrt{2\,\epsilon_1}}\ ,
\label{a.4}
\eeq
 \beq
 \Psi^-_F(\vec x,\vec p_2) =
e^{ip_2z}\,
 \hat D_2\,F^-(\vec x,\vec
p_2)\,
 \frac{u(\vec p_2)}{\sqrt{2\,\epsilon_2}}\ ,
\label{a.5}
\eeq
 where 
 \beqn
 \hat D_{1,2} &=&
1- i\,\frac{\widehat{\vec\alpha}\cdot \vec\nabla}
 {2\,\epsilon_{1,2}} -
\frac{\widehat{\vec\alpha}(
 \vec p_{1,2}-\vec n\,p_{1,2})}{2\,\epsilon_{1,2}}\ ;\\
\vec n &=& \frac{\vec k}{k}\ ;\nonumber\\
z &=& \vec n\cdot\vec x\ ;\nonumber\\
p_{1,2} &=& |\vec p_{1,2}|\ .\nonumber
\label{a.6}
\eeqn

In this case the functions $F(\vec x,\vec p)$ and $f(\vec x,\vec p)$
are related as,
\beq
F(\vec x,\vec p)={\rm exp}(i\vec p\vec x - ipz)\,f(\vec x,\vec p)\ .
\label{a.6a}
\eeq
Therefore, $F(\vec x,\vec p)=F^{\pm}(\vec x,\vec p)$ 
has to satisfy the equation,
\beq
\left(\Delta + 2\,i\,p\,\frac{d}{dz} -
2\,\epsilon\,U(\vec x)\right)\,F(\vec x,\vec p)\ .
\label{a.6b}
\eeq
The characteristic longitudinal distances in the
problem under consideration $x_L\sim \epsilon/m^2$
are much longer than the typical transverse distances $x_T\sim 1/m$
\cite{ll}. Therefore, in the Laplacian $\Delta=d^2/dz^2 + (d/d\vec x)^2$
one can drop the first term $d^2/dz^2$. Then (\ref{a.6b})
takes the form of the two-dimensional Schr\"odinger equation,
 \beq
 i\,\frac{{\it d}}{{\it d}z}\,
 F(\vec x,\vec p) =
\left[-\frac{\Delta_T}{2\,p} +
 U(\vec x)\right]\, F(\vec x,\vec p)\ ,
\label{a.7}
\eeq
where $p=|\vec p|$.
 We define $F^{\pm}$ in accordance with the asymptotic behavior,
\beq
 F^+(\vec x,\vec p_1)\Bigr|_{z\to z_-=-\infty}
 \rightarrow e^{i\,\vec
p_{1T}\,\vec r}
\label{a.8}
\eeq
 \beq
 F^-(\vec x,\vec p_2)\Bigr|_{z\to z_+=+\infty}
 \rightarrow
e^{i\,\vec p_{2T}\,\vec r}\ .
\label{a.9}
\eeq
 Here we introduced new notations for transverse, $\vec r\equiv
\vec x_T$, and longitudinal, $z\equiv x_L$, coordinates.

It follows from (\ref{a.7})
- (\ref{a.9}) that these functions can be represented
 in the form,
\beq
 F^+(\vec x,\vec p_1) =
 \int d^3r_1\,G(z,\vec r;z_-,\vec r_1|\vec
p_1)\,
 e^{i\,\vec p_{1T}\,\vec r_1}\ ,
\label{a.10}
\eeq
 \beq
 {F^-}^*(\vec x,\vec p_2) =
 \int d^3r_2\,G(z_+,\vec
r_2;z,\vec r|\vec p_2)\,
 e^{- i\,\vec p_{2T}\,\vec r_2}\ ,
\label{a.11}
\eeq
 where $G(z_2,\vec r_2;z_1,\vec r_1|\vec p)$ is the retarded
 Green
function corresponding to Eq.~(\ref{a.7}),
 \beq
 \left[i\,\frac{{\it
d}}{{\it d}z_2} +
 \frac{\Delta_2}{2\,p} -
 U(z_2,\vec r_2)\right]\,
G(z_2,\vec r_2;z_1,\vec r_1|\vec p) =
 i\,\delta(z_2-z_1)\,\delta(\vec
r_2-\vec r_1)
\label{a.12}
\eeq
 and satisfying the conditions,
 \beqn
 G(z_2,\vec r_2;z_1,\vec
r_1|\vec p)\Bigr|_{z_1=z_2} &=&
 \delta(\vec r_2-\vec r_1)\nonumber\\
G(z_2,\vec r_2;z_1,\vec r_1|\vec p)\Bigr|_{z_1>z_2}= 0\ .
\label{a.13}
\eeqn

 It is convenient to chose the axis $z$ along the momentum of
the
 radiated photon. Then
 \beqn
 \vec p_{1T} &=& - \frac{\vec
k_T}{\alpha}\ ,\nonumber\\
 \vec p_{2T} &=& \vec p_T -
\frac{1-\alpha}{\alpha}\, \vec k_T\ ,
\label{a.14}
\eeqn
 where $\vec k_T$ and $\vec p_T$ are the transverse components
 of
the photon and final electron momenta relative to the direction of
 the initial
electron; $\alpha$ is the fraction of the
 light-cone momentum of the initial
electron carried by the photon.

 We arrive at the following expression for
the radiation amplitude
 (\ref{a.2}),
 \beqn
 M_{fi}^T &=&
\frac{1}{2\,p\,(1-\alpha)}\,
 \int d^2r_1\,d^2r_2\,d^2r\,dz\,
 {\rm exp}(-
i\,\vec p_{2\,T}\,\vec r_2)\,
 G(z_+,r_2;z,r|\vec p_2)\nonumber\\
 &&{\rm
exp}(i\,q_{min}\,z)\,
 \widehat \Gamma\,
 G(z,r;z_-,r_1|\vec p_1)\,
 {\rm
exp}(i\,\vec p_{1\,T}\,\vec r_1)\ ,
\label{a.15}
\eeqn
where
\beq
q_{min}=\frac{\alpha\,m_q^2}{2(1-\alpha)E_q}+
\frac{m^2_{\gamma^*}}{2\alpha E_q}\ ,
\label{a.15a}
\eeq
and $E_q$, $m_q$ are the energy and the mass of the projectile quark.
In the approximation considered in this paper when the fluctuation
time substantially exceeds the interaction time, $q_{min}\ll 1/R_A$
and can be neglected.
 
The vertex function in (\ref{a.15}) reads,
 \beqn
 \widehat\Gamma &=&
\sqrt{1-\alpha}\,u^*(\vec p_2)\,\hat D^*_2\,
 \widehat{\vec\alpha}\cdot \vec {e^*}\,\hat D_1\,u(p_1)
 \nonumber\\
&=& \chi^{\dagger}_2\,\left[i\,m\,\alpha\,
 (\vec n\times\vec\sigma)\cdot \vec {e^*} +
\alpha\,(\sigma\times\vec\nabla_T)\cdot
 \vec {e^*} -
i\,(2-\alpha)\,\vec\nabla_T\cdot \vec {e^*}\right]\,\chi_1\ .
\label{a.16}
\eeqn
The operator $\vec\nabla_T=d/d\vec r$ acts to the right.
$\chi_{1,2}$ are the two-component spinors of the initial and final
electrons.

 In the case of a composite target the potential has to be summed over
the constituents,
\beq
U(\vec r,z)=
 \sum_i\,U_0(\vec r-\vec r_i, z-z_i)
\label{a.17}
\eeq
and the bremsstrahlung cross section
should be averaged over the positions $(\vec r_i,z_i)$ of the
scattering centres. 

The averaged matrix element squared 
takes the form,
 \beqn
\left\la\left|M^T_{fi}\right|^2\right\ra &=&
 2\,{\rm Re}\,\int\limits_{-\infty}^{\infty}dz_1
\int\limits_{z_1}^{\infty}dz_2\int
 d^2r_1\,d^2r_1'\,d^2r_2\,d^2r_2'\,d^2r\,d^2r'\,
d^2\rho\,d^2\rho'\nonumber\\
 &\times& {\rm
exp}\left[i\,\vec p_{2\,T}\,(\vec{r_2'}-\vec r_2) -
i\,\vec p_{1\,T}\,(\vec{r_1'}-\vec r_1)
- iq_{min}(z_2-z_1)\right]\,
\nonumber\\
&\times& \left\la G(z_+,\vec
r_2;z_2,\vec\rho\,|\,p_2)\,
G^*(z_+,\vec{r'_2};z_2,\vec{r'}
\,|\,p_2)\right\ra\nonumber\\
 &\times& \widehat\Gamma^{'*}\,
 \left\la G(z_2,\vec\rho;z_1,\vec
r\,|\,p_2)\,
G^*(z_2,\vec {r'};z_1,\vec{\rho'}\,|\,p_1)\right\ra\nonumber\\
 &\times&
\widehat{\Gamma}\,
 \left\la G(z_1,\vec r;z_-,\vec r_1\,|\,p_1)\,
G^*(z_1,\vec {\rho'};z_-,\vec r_1'\,|\,p_1)\right\ra\ ,
\label{a.18}
\eeqn
 where ${\widehat\Gamma}'$ differs from $\widehat\Gamma$ in
(\ref{a.16}) by the replacement $$ \vec\nabla=\frac{d}{d\vec r}
\Rightarrow\vec\nabla^{~\prime}=\frac{d}{d\vec
r^{~\prime}}\, .$$

The following consideration is based on the representation of 
the Green function $G$ in the form of a continuous integral \cite{feynman},
\beq
G(z_2,\vec r_2;z_1,\vec r_1\,|\,p) =
\int {\cal D}\vec r(z)\,
{\rm exp}\left\{\frac{ip}{2}\, \int\limits_{z_1}^{z_2}dz\,
\left(\frac{d\vec r(z)}{dz}\right)^2 -
i\,\int\limits_{z_1}^{z_2}dz\,U\bigl(\vec r(z),z\bigr)\right\}\ ,
\label{a.19}
\eeq
where
$$\vec r(z_1)=\vec r_1,~~\vec r(z_2)=\vec r_2\, ,$$
and the relation
\beq
\int\limits_{z_1}^{z_2}dz\,\sum\limits_i
U_0\bigl(\vec r(z)- \vec r_i, z-z_i\bigr) = 
\sum\limits_i \chi\bigl(\vec r(z_i)-\vec r_i\bigr)\,
\Theta(z_2-z_i)\,\Theta(z_i-z_1)\ ,
\label{a.20}
\eeq
where $\chi(\vec r) = \int_{-\infty}^{\infty}dz\,U_0(\vec r,z)$.

The mean value of the eikonal exponential is,
\beqn
&&\left<\exp\left\{i\sum_{i}
\left[\chi\left(\vec r(z_i)\right)-\chi\left(\vec
r^{~\prime}(z_i)\right)\right]\Theta(z_2-z_i)\Theta(z_i-z_1)
\right\}\right> =\nonumber\\
&&\exp\left\{-\frac{1}{2}\int\limits_{z_1}^{z_2}dz\,n(z,\vec b)
\sigma\left[\vec r(z)-\vec r^{~\prime}(z)\right]\right\}\, , 
\label{a21}
\eeqn
where
\beq
\sigma(\vec r-\vec r^{~\prime})=2\int d^2\rho\left[1-\exp\left(
i\chi(\vec r-\vec\rho)-i\chi(\vec r^{~\prime}-\vec\rho) \right)\right]\, ,
\label{a.22}
\eeq
and $n(z,\vec b)$ is the density of scattering centres.

Using these relations and performing integration 
by parts in (\ref{a.18}),
\beqn
\frac{d\sigma^T}{d({\rm ln}\alpha) d^2p_Td^2k_T} &=&
\frac{\alpha_{em}}{(2\pi)^4\,4\,p^2\,(1-\alpha)^2}
2Re\int\limits_{-\infty}^{\infty}dz_1\int\limits_{z_1}^{\infty}dz_2
\int d^2b~d^2\rho_1d^2\rho_2\nonumber\\
&\times& {\rm exp}\left[i\alpha\,\vec p_{2T}\,\vec\rho_{2}-
i\alpha\,\vec p_{1T}\,\vec \rho_{1} -
\int\limits_{z_2}^{\infty}dz\,V(z,\vec\rho_{2}) -
\int\limits_{-\infty}^{z_1}dz\,V(z,\vec\rho_{1})\right]\nonumber\\
&\times&\widehat{\gamma}_2\,\widehat{\gamma}_{1}^{\ast}\,
W(z_2,\vec\rho_2;z_1,\vec\rho_1\,|\,p)\, .
\label{a.23}
\eeqn
The variables in this equation are related to those in (\ref{a.18}) as,
\beqn
\vec\rho_1 &=& \frac{\vec r_1\!' - \vec r_1}{\alpha}\nonumber\\
\vec\rho_2 &=& \frac{\vec r_2\!' - \vec r_2}{\alpha}\nonumber\\
\vec b &=& {1\over 2}(\vec r_1\!' + \vec r_1)\nonumber\ .
\label{a.23a}
\eeqn
Other variables in (\ref{a.18}) are integrated explicitly.

Matrices $\widehat{\gamma}$ are related to $\widehat{\Gamma}$ in
(\ref{a.16}) by replacement $m\Rightarrow \alpha m$ and $d/d\vec r \Rightarrow
d/d\vec\rho$.

Absorptive potential $V$ in (\ref{a.23}) reads,
$$V(z,\vec\rho)=n(z,\vec b)\frac{\sigma}{2}(\alpha\cdot\vec\rho)\, ,$$
and $W$ is the solution of either of the equations,
\beq
\frac{\partial}{\partial
z_2}W(z_2,\vec\rho_2;z_1,\vec\rho_1\,|\,p)=
\frac{i\left[\Delta(\vec\rho_2)-
\varepsilon^2 \right]}{2\alpha(1-\alpha)p}\,
W(z_2,\vec\rho_2;z_1,\vec\rho_1\,|\,p) - 
V(\vec\rho_2,z_2)\,W(z_2,\vec\rho_2;z_1,\vec\rho_1\,|\,p)\, ,
\label{a.24}
\eeq
\beq
-\frac{\partial}{\partial z_1}W(z_2,\vec\rho_2;z_1,\vec\rho_1\,|\,p)=
\frac{i\left[\Delta(\vec\rho_1)-
\varepsilon^2 \right]}{2\alpha(1-\alpha)p}\,
W(z_2,\vec\rho_2;z_1,\vec\rho_1\,|\,p) - 
V(\vec\rho_1,z_1)\,W(z_2,\vec\rho_2;z_1,\vec\rho_1\,|\,p)\, ,
\label{a.25}
\eeq
with the boundary condition
\beq
W(z_2,\vec\rho_2;z_1,\vec\rho_1\,|\,
p)\biggl\vert_{z_2=z_1}=\delta(\vec\rho_2-\vec\rho_1)\, .
\label{a.26}
\eeq
Using these equations and the relation,
\beq
\left[\Delta(\vec\rho)-\varepsilon^2
\right]K_0\left(
\varepsilon\left\vert\vec\rho~\right\vert\,\right)=
-2\pi\delta(\vec\rho)
\label{a.27}
\eeq
simple but cumbersome calculations lead to a new form for Eq.~(\ref{a.23}),
\beqn
\frac{d\sigma^T}{d({\rm ln}\alpha)\,d^2p_T\,d^2 k_T} &=&
\frac{\alpha^2}{(2\pi)^4} \Biggl\{ {\rm Re}\int\limits_{-\infty}^{\infty}dz
\int d^2b~d^2\rho_1d^2\rho_2~d^2 \rho\nonumber\\
&\times&{\rm exp}\left[i\alpha\,\vec
p_{2T}\,\vec\rho_{2}- i\alpha\,\vec p_{1T}\,\vec
\rho_{1}-\int\limits_{z}^{\infty}dz'\,V(z^{\prime},\vec\rho_{2})
-\int\limits_{-\infty}^{z}dz'\,V(z^{\prime},\vec\rho_{1})\right]
\nonumber\\
&\times&\Psi_{T}^{\dagger}(\vec\rho_{2}-\vec\rho)\,\Bigl[2\,V(z,\vec\rho)-V(z,
\vec\rho_1)-V(z,\vec\rho_2) \Bigr]\,\Psi_{T}(\vec\rho_{1}-\vec\rho)
\nonumber\\
&-&2{\rm Re}\int\limits_{-\infty}^{\infty}dz_1\int\limits_{z_1}^{\infty}dz_2
\int d^2b~d^2\rho_1d^2\rho_2d^2\rho_1^{\prime}d^2\rho_2^{\prime}
\nonumber\\
&\times&{\rm exp}\left[i\alpha\,\vec p_{2T}\,\vec\rho_{2}-
i\alpha\,\vec p_{1T}\,\vec \rho_{1}-\int\limits_{z_2}^{\infty}dz\,V(z,\vec\rho_{2})
-\int\limits_{-\infty}^{z_1}dz\,V(z,\vec\rho_{1})\right]
\nonumber\\
&\times&\Psi_{T}^{\dagger}(\vec\rho_{2}-\vec\rho_{2}^{~\prime})
\,\Bigl[V(z_2,\vec\rho_2)-V(z_2,\vec\rho_2^{~\prime})\Bigr]\,
W(z_2,\vec\rho_2';z_1,\vec\rho_1'\,|\,p)\nonumber\\
&\times&
\Bigl[V(z_1,\vec\rho_1)-V(z_1,\vec\rho_1^{~\prime})\Bigr]\,
\Psi_{T}(\vec\rho_{1}-\vec\rho_{1}^{~\prime})\Biggr\}\, ,
\label{a.28}
\eeqn
where
\beq
\Psi_T(\vec\rho)=\frac{\sqrt{\alpha_{em}}}{2\pi}~\widehat{\Gamma}~
K_0(\varepsilon\rho)\, .
\label{a.28a}
\eeq

In the ultrarelativistic limit $(p\to\infty)$ we have
\beq
W(z_2,\vec\rho_2;z_1,\vec\rho_1\,|\,
\infty)=\delta(\vec\rho_2-\vec\rho_1)\,{\rm exp}
\left[-\int\limits_{z_1}^{z_2}dz\,V(z,\vec\rho_{2})\right]\ .
\label{a.29}
\eeq
The integrations over $z$, $z_1$, $z_2$ in (\ref{a.28})
can be performed analytically, and we arrive at the expression
\beqn
\frac{d\sigma^T}{d({\rm ln}\alpha)\, d^2p_T\,d^2k_T} &=& 
\frac{\alpha^2}{(2\pi)^4}\, \int d^2r_1\,d^2r_2\,d^2r\,
{\rm exp}\left[i\,\alpha\,\vec r\,(\vec p_{T}+\vec k_{T})
+i\,(\vec r_1-\vec r_2)\,\vec k_{T}\right]
\nonumber\\ &\times&
\psi_{T}(\vec r_{1})\,\psi_{T}^{\ast}(\vec r_{2})\,
\Sigma_{\gamma}(\vec r,\vec r_1,\vec r_2,\alpha)\ ,
\label{a.30}
\eeqn
where
\beq
\Sigma_{\gamma}(\vec r,\vec r_1,\vec r_2,\alpha)=
\Sigma\biggl(\alpha(\vec r+\vec r_1)\biggl)+
\Sigma\biggl(\alpha(\vec r - \vec r_2)\biggl) -
\Sigma(\alpha\vec r)-
\Sigma\biggl(
\alpha(\vec r + \vec r_1-\vec r_2)\biggl)\, , 
\label{a.30a}
\eeq
and
\beq
\Sigma(\rho)=\int d^2b\,\left\{1-
{\rm exp}\left[-\frac{\sigma(\rho)}{2}\,
T(b)\right]\right\}\, .
\label{a.31}
\eeq

The derivation of the correspondent expressions for gluon bremsstrahlung
is done analogously. We skip the details and present only the results.
\beq
\Sigma_g(\vec r,\vec r_1,\vec r_2,\alpha)=
\Sigma_1(\vec r,\vec r_1,\vec r_2,\alpha)+
\Sigma_2(\vec r,\vec r_1,\vec r_2,\alpha)-
\Sigma_3(\vec r,\vec r_1,\vec r_2,\alpha)-
\Sigma_4(\vec r,\vec r_1,\vec r_2,\alpha)\ ,
\label{a.32}
\eeq
where
\beqn
\Sigma_i(\vec r,\vec r_1,\vec r_2,\alpha)&=&
\int d^2b\,\left\{1- {\rm exp}
\left[-{1\over2}\,\sigma_i(\vec r,\vec r_1,\vec r_2,\alpha)\,
T(b)\right]\right\}\ ;
\label{a.33}\\
\sigma_1(\vec r,\vec r_1,\vec r_2,\alpha)&=&
{9\over8}\left[\sigma\Bigl(\vec r+(1-\alpha)\vec r_2\Bigr) + 
\sigma(\vec r_1)\right] -
{1\over8}\sigma(\vec r+\alpha \vec r_1)\ ;\\
\sigma_2(\vec r,\vec r_1,\vec r_2,\alpha)&=&
{9\over8}\left[\sigma\Bigl(\vec r-(1-\alpha)\vec r_2\Bigr) +
\sigma(\vec r_2)\right] -
{1\over8}\sigma(\vec r+\alpha \vec r_2)\ ;\\
\sigma_3(\vec r,\vec r_1,\vec r_2,\alpha)&=&
\sigma(\alpha\vec r)\ ;\\
\sigma_4(\vec r,\vec r_1,\vec r_2,\alpha)&=&
\sigma\Bigl(\vec r-\alpha(\vec r_1-\vec r_2)\Bigr) +
{9\over4}\sigma\Bigl(\vec r + (1-\alpha)(\vec r_1-\vec r_2)\Bigr)+
{9\over8}\Bigl[\sigma(\vec r_1)+\sigma(\vec r_2)
\nonumber\\&-&
\left.\sigma\Bigl(\vec r+(1-\alpha\vec r_1+\alpha\vec r_2\Bigr) -
\sigma\Bigl(\vec r-(1-\alpha\vec r_2+\alpha\vec r_1\Bigr)\right]\ .
\eeqn

This expression simplifies and gets the form of (\ref{4.8})
if one integrates in (\ref{a.30}) over transverse momentum
$p_T$ of the quark. Note that the last cross section
$\sigma_4(\vec r,\vec r_1,\vec r_2,\alpha)$ is the total
cross section for a colorless system of two gluons, quark 
1and antiquark interacting with a nucleon (compare with
(\ref{2.8})). Here $\vec r_1$ and $\vec r_2$ are the transverse 
separations inside the $qg$ and $\bar qg$ pairs and $\vec r$ is
the transverse distance between the centers of gravity of these 
pairs.

\section{}
\setcounter{equation}{0}

In order to calculate Eqs.~(\ref{3.2}) - (\ref{3.3}) in 
the dipole approximation $\sigma_{q\bar q}=C\:r^2$, we
need to evaluate integrals of two types:
\beqn
I_1&\!=\!&\frac{1}{(2\pi)^2}\,\int d^2\, r_1d^2 r_2\,\exp \left[i\vec
k_T\:(\vec r_1-\vec r_2)\right]\nonumber \\
&&\times\, K_0(\varepsilon r_1)K_0(\varepsilon r_2)\,
\exp\left\{-\:\frac{1}{4}\left(fr_{1}^{2}+hr_{2}^{2}-2g\vec r_1\vec
r_2\right) \right\}\label{e1}\,;
\label{b.1}
\eeqn
and
\beqn
I_2&\!=\!&\frac{1}{(2\pi)^2}\,\int d^2 r_1d^2 r_2\,
\exp\left[i\vec k_T\:(\vec
r_1-\vec r_2)\right] \nonumber\\
&&\times \,\frac{(\vec r_1\vec r_2)}{r_1r_2}\,K_1(\varepsilon
r_1)K_1(\varepsilon r_2)\,
\exp\left\{-\,\frac{1}{4}\left(fr_{1}^{2}+hr_{2}^{2}-2g\vec r_1\vec
r_2\right)\right\}\label{e2}\,.
\label{b.2}
\eeqn
Here we use the notation,
\beq
\frac{\sigma_{\bar qq}(\rho)}{2}\;T(b)\:=\;\frac{1}{4}
\left(fr_{1}^{2}+hr_{2}^{2}-2g\vec r_1\vec r_2\right)\, .
\label{b.3}
\eeq

We use the 
integral representation for the modified Bessel functions,
which reads
\beq
K_0\,(\varepsilon r)\,=\:\frac{1}{2}\,\int\limits_{0}^{\infty}\frac{dt}{t}\,
\exp\left\{-t-\frac{\varepsilon^2r^2}{4t}  \right\} \,;
\end{equation}
\begin{equation}
\label{e5}
\frac{1}{\varepsilon\,r}\,K_1(\varepsilon\,r)\:=\:\frac{1}{4}\,\int\limits_{0}^
{\infty}\frac{dt}{t^2}\,
\exp\left\{-t-\frac{\varepsilon^2r^2}{4t}  \right\}\, .
\label{b.4}
\eeq

After substitution of (\ref{b.4}) and (\ref{b.5}) into (\ref{b.1}) and
(\ref{b.2}) and making use of the following obvious relations,
\beqn
I_3&\!=\!&\frac{1}{4\,(2\pi)^2}\,\int d^2 r_1d^2 r_2\,\exp \left\{i\vec
k_T(\vec r_1-\vec r_2)\right. \nonumber \\
&&-\,\left.\frac{1}{4}\left(a\,r_{1}^{2}+c\,r_{2}^{2}-2b\,\vec r_1\vec
r_2\right) \right\}\label{e6}\\
&=&\frac{1}{(ac-b^2)}\,
\exp\left\{-\,\frac{k_{T}^{2}\,(a+c-2b)}{(ac-b^2)} \right\}\,; \nonumber
\label{b.5}
\eeqn
\beqn
I_4&\!=\!&\frac{1}{16\,(2\pi)^2}\,\int d^2 r_1d^2 r_2\,(\vec r_1\vec r_2)\,
\exp\left\{i\vec k_T(\vec r_1-\vec r_2)\right.\nonumber\\
&& -\,\left.\frac{1}{4}\left(a\,r_{1}^{2}+c\,r_{2}^{2}-2b\,\vec r_1\vec
r_2\right) \right\}\nonumber\\
&=&\left[\frac{1}{(ac-b^2)^2}-\frac{b\,k_{T}^{2}(a+c-2b)}{(ac-b^2)^3}\right]
\times\exp\left\{-\,\frac{k_{T}^{2}(a+c-2b)}{ac-b^2} \right\}\label{e7}\,.
\label{b.6}
\eeqn
one arrives at,
\beqn
I_1&\!=\!&\int\:\frac{dt}{t}\,\frac{du}{u}\,\exp(-u-t)\;\;I_3 
\label{b.7}\,
,\\
I_2&\!=\!&\varepsilon^2\int\frac{dt\,du}{t^2\: u^2}\,
\exp(-u-t)\;\,I_4\nonumber;
\label{b.8}
\eeqn
where
\beq
a=\frac{\varepsilon^2}{t}+f,\quad 
c=\frac{\varepsilon^2}{u}+h,\quad 
b=g\,.
\label{b.9}
\eeq

Thus, for the  general case in addition to 
the integration over the impact parameter one 
has to evaluate  numericaly a two-dimensional integral over 
$dt$ and $du$.

The situation is simplified in the case of photon bremsstrahlung, when
integration for the three exponentials in (\ref{3.2}) correspond to
the following values of the
parameters, respectively,
\beq
\begin{array}{cccc}
f=g=0\, ,& h= &2c\alpha^2\,T(b)\, ; \\
h=g=0\, ,& f= &2c\alpha^2\,T(b)\, ; \\
f\,=\,h\,=\,g&\,=\, &2c\alpha^2\,T(b)\,. \\
\end{array}
\label{b.10}
\eeq
In this case Eqs.~(\ref{b.5}) and  (\ref{b.6}) are reduced 
to one-dimensional integrals.

\end{document}